\documentclass[12pt]{article}

\catcode`\@=11

\global\arraycolsep=2pt
\usepackage{amsbsy,amssymb,latexsym,amsfonts,amsmath}
\usepackage{graphicx,color}

\begin{document}
\begin{flushright}
\parbox{4.2cm}
{RUP-18-11}
\end{flushright}

\vspace*{0.7cm}

\begin{center}
{ \Large On the realization of impossible anomalies}
\vspace*{1.5cm}\\
{Yu Nakayama}
\end{center}
\vspace*{1.0cm}
\begin{center}

Department of Physics, Rikkyo University, Toshima, Tokyo 171-8501, Japan

\vspace{3.8cm}
\end{center}

\begin{abstract}
The Wess-Zumino consistency condition allows more exotic forms of anomalies than those we usually encounter. For example in two-dimensional conformal field theories in the curved background with space-time dependent coupling constant $\lambda^i(x)$, a $U(1)$ current could possess anomalous divergence of the form $D^\mu J_\mu = \tilde{c} R + \chi_{ij} \partial^\mu  \lambda^i \partial_\mu\lambda_j + \tilde{\chi}_{ij} \epsilon^{\mu\nu} \partial_\mu \lambda^i \partial_\nu \lambda^j + \cdots $. Another example is the CP odd Pontryagin density in four-dimensional Weyl anomaly.
We could, however, argue that they are impossible in conformal field theories because they cannot correspond to any (unregularized) conformally invariant correlation functions. We find that this no-go argument may be a red herring. We show that some of these impossible anomalies avoid the no-go argument because they are not primary operators, and the others circumvent it because they are realized as semi-local terms as is the case with the conformally invariant Green-Schwartz mechanism and in the higher dimensional analogue of Liouville or linear dilaton theory.

\end{abstract}

\thispagestyle{empty} 

\setcounter{page}{0}

\newpage

\section{Introduction}
Anomalies\footnote{In this paper, we define the anomaly as $c-$number violation of the conservation law under the presence of the non-trivial background field that cannot be removed by adding local $c$-number counterterms. }
 in quantum field theories are constrained from their algebraic structures given by the Wess-Zumino consistency condition \cite{Wess:1971yu}, which demands that the symmetry transformation is integrable. One elegant solution of the Wess-Zumino consistency condition is given by solving the descent equation and relating it to the higher-dimensional anomaly polynomial \cite{Manes:1985df}. It has a beautiful geometric realization as well as physical realization by the so-called symmetry-protected-topological phases of matter. In addition, the Wess-Zumino consistency condition for the Weyl anomaly together with the local renormalization group gives non-trivial constraint on the renormalization group flow, and has attracted a lot of attentions over years \cite{Osborn:1991gm}\cite{Nakayama:2013wda}\cite{Jack:2013sha}.
 Moreover the effective field theory realization of the Weyl anomaly is a starting point of the proof of the $a$-theorem \cite{Komargodski:2011vj} and discussions on the equivalence between scale invariance and conformal invariance in four-dimensions \cite{Luty:2012ww}\cite{Dymarsky:2014zja}. See e.g. \cite{Nakayama:2013is} for a review.

We, however, note that the there are more solutions to the Wess-Zumino consistency conditions than those obtained from the anomaly polynomials. For instance, let us consider four-dimensional field theories with the anomalous conservation of the $U(1)$ current under the presence of (the other) background $U(1)$ gauge fields $A_\mu$ with the field strength $F_{\mu\nu}$. In addition to the conventional Adler-Bell-Jackiw anomaly \cite{Adler:1969gk}\cite{Bell:1969ts}
\begin{align}
\partial^\mu J_\mu =  t_1 \epsilon^{\mu\nu\rho\sigma} F_{\mu\nu} F_{\rho\sigma}
\end{align}
we could have the additional (non-conventional) structure of the form
\begin{align}
\partial^\mu J_\mu =  t_2 F_{\mu\nu} F^{\mu\nu} \label{exist} \ .
\end{align}
We can easily see that the both forms of anomaly are allowed by the Wess-Zumino consistency conditions although the usual descent formalism from the higher dimensional anomaly polynomial does not give the second term.\footnote{In contrast, in the case of the non-Abelian anomaly (i.e. the violation of $D^\mu J^a_\mu = 0$), the Wess-Zumino consistency condition is strong enough to fix the form of the anomaly in the conventional form up to an overall factor.}  
Is there any theoretical principle that the second form does not appear possibly in CP violating theories? Of course, with a given Lagrangian theory, one may argue its absence by perturbative discussions similarly to the Adler-Bardeen theorem \cite{Adler:1969er}, and the usual argument goes as follows. Suppose the theory under consideration is defined as an asymptotic free theory. Then we may use t'Hooft anomaly matching argument to evaluate the anomaly in the ultraviolet theory, but it is just fixed by the one-loop diagram. Beyond the perturbation theory or especially in the non-Lagrangian theories, we might wonder what would be the fundamental obstructions.\footnote{In this particular case, one may resort to the quantization of the $U(1)$ charge and the compactness of the $U(1)$ gauge symmetry to discard the possibility \eqref{exist}, but we do not always have such  and argument and the necessity of the quantization could be questioned.}

If there were such anomalies, we might achieve more intrinsic classifications of (possibly CP violating) quantum field theories. They might give further constraint on the renormalization group from the analogue of the 't Hooft anomaly matching. For example, if such theories exist, they cannot be gapped.
They might have particle physics applications as we had in the $\pi_0 \to 2\gamma$ decay in the ordinary chiral anomaly.

On the other hand, as we will discuss in the main part of the paper, there is an argument that these anomalies cannot be realized in conformal field theories. The main point of the argument is that if there existed such terms in the anomaly, there must exist a corresponding three-point function $\langle J_\mu J_\nu J_\rho \rangle$ which reduces to \eqref{exist} after taking the divergence. The analysis of the conformal symmetry, however, tells that such a (unregularized) three-point function supported at the non-coincident point does not exist and hence it is impossible. The argument sounds convincing but slightly mysterious. Certainly the anomalous current conservation law itself is Weyl invariant and there is no violation of the Wess-Zumino consistency condition either for the $U(1)$ symmetry or the Weyl transformation. Then what is the underlying conceptual reason why these terms are not allowed? In other words, how can we evade the totalitarian principle of Gell-Mann: ``Everything not forbidden is compulsory."?

In this paper, we call these anomalies as ``impossible anomalies" and study the properties and possible realizations. On one hand, the impossible anomalies look perfectly healthy and consistent but on the other hand, it seems that they are not compatible with the conformal symmetry in their actual realizations. Our goal it to try to resolve this dilemma in two different ways. In both cases, we find that the no-go argument above may be a red herring. In the first case discussed in section 2, we show that the no-go argument can be avoided by realizing the current operators may  not be primary operators. In the second case discussed in section 3, we show that the no-go argument can be circumvented by realizing the anomalous correlation functions can be only semi-local as is the case with the conformally invariant Green-Schwartz mechanism. We conclude the paper with discussions in section 4.

\section{Realizing impossible anomalies from descendants}
In this section, we study the first mechanism to avoid the no-go argument for impossible anomalies. The main idea is to relax the condition that the current operator is a primary operator. In fact, the idea itself is ubiquitous and quite commonly observed in two-dimensional conformal field theories, so we begin with our analysis in two-dimensions.

We study a two-dimensional conformal field theory with conserved $U(1)$ currents $J^V_\mu$ and $J^A_\mu$. Here superscript $V$ stands for the vector current whose left-mover is $J$ and whose right mover is $\bar{J}$, and $A$ stands for the axial current  whose left-mover is $J$ and whose right mover is $-\bar{J}$.
Let us put the theory in the curved space-time with the Ricci scalar given by  $R$. The $U(1)$ current may be anomalous in the curved background, and we consider the possible anomaly of the form
\begin{align}
D^\mu J^V_\mu = a R + \cdots, \label{cano}
\end{align}
where 
 $\cdots$ means the other anomaly terms that we will not discuss for now.

This anomaly term is allowed in the sense of the Wess-Zumino consistency condition. Indeed, the Wess-Zumino consistency condition does not say much about the possible form of the $U(1)$ current anomaly. The commutative nature of the $U(1)$ anomaly
\begin{align}
[\delta_{\lambda_1(x)},\delta_{\lambda_2(y)}] = 0
\end{align}
implies that if the partition function $Z[A_\mu(x)]$ shows the anomalous variation$
\mathcal{A}[\lambda(x),A_\mu(x)] = \delta_{\lambda(x)} \log Z[A_\mu(x)]$ under the gauge transformation $\delta A_{\mu}(x) = \partial_\mu \lambda(x)$ for the background gauge field $A_{\mu}(x)$ that couples with $J_\mu^V$, it must satisfy the algebraic constraint
\begin{align}
\delta_{\lambda_1(x)} \mathcal{A}[\lambda_2(x)] = \delta_{\lambda_2(x)} \mathcal{A}[\lambda_1(x)] \ .
\end{align}
This condition, however, is trivially true when the anomalous variation is gauge invariant as in \eqref{cano}
\begin{align}
\mathcal{A}[\lambda(x)] = \int d^2x\sqrt{g} \lambda(x) a R \ . 
\end{align}
Therefore the anomaly of the form \eqref{cano} is integrable and perfectly healthy in this sense.

However, a closer inspection might indicate that such an anomaly cannot exist in conformal field theories from the following  argument. Suppose the anomaly is realized as in \eqref{cano}. Then it must be visible from the study of the two-point functions of the energy-momentum tensor $T_{\mu\nu}$ and the current $J_\mu$. 
More precisely, it should be related to the two-point function
\begin{align}
\langle T(z) J(0) \rangle = \frac{4a}{z^3} \ ,
\end{align}
where we have introduced the complex coordinate $z= x_1+ix_2$ and holomorphic tensors $T= T_{zz}$ and $J= J_{z}$ as usual in two-dimensional conformal field theory. Indeed the divergence  gives the anomalous conservation 
\begin{align}
\bar{\partial} \langle \tilde{T}(z) J(0) \rangle =  4\pi a \partial_z^2 \delta^{(2)}(z,\bar{z})  
\end{align}
from the formula $\bar{\partial}\frac{1}{z} = 2\pi \delta^{(2)}(z,\bar{z})$, 
 which is equivalent to \eqref{cano}.

On the other hand, if we assume that $T$ and $J$ are (quasi-)primary operators, we immediately realize that the conformal invariance demands that the two-point functions between primary operators of  different twists $\Delta\pm J$ vanish. Since $T$ and $J$ have different twists, we conclude $\alpha =0$ in conformal field theories. As a matter of fact, the anomaly equation \eqref{cano} itself may not look Weyl invariant from the beginning because the Ricci scalar  has the non-trivial Weyl transformation $R \to e^{-2\sigma} (R - 2\Box \sigma)$, and does not seem to make sense in conformal field theories.  We will come back to this point later after showing how to circumvent this no-go argument.

Nevertheless, we actually know that such anomalies do exist. For example, if we study the string worldsheet theory, the ghost number conservation is anomalous and indeed it has the same form as in \eqref{cano} (see e.g. \cite{Polchinski:1998rq}]). The above no-go argument is avoided because the ghost number current is not a primary operator. In unitary conformal field theories, conserved current operators are necessarily primary operators, but it is not the case here.

Actually, the situation is more generic. Let us consider any two-dimensional conformal field theories with $U(1)$ current algebra with the standard operator product expansion (OPE):
\begin{align}
\langle J(z) J(0) \rangle = \frac{k}{2z^2} \cr
\langle T(z) T(0) \rangle = \frac{c}{2z^4} \cr
\langle T(z) J(0) \rangle = 0
\end{align}
For simplicity, we assume that the theory is left-right symmetric so that the right mover with $(\bar{T}, \bar{J})$ has the same OPE with $z$ replaced by $\bar{z}$.

Let us now define the twisted energy-momentum operator by
\begin{align}
\tilde{T} &= T + \alpha \partial J \cr
\tilde{\bar{T}} &= \bar{T} + \alpha \bar{\partial} \bar{J} \ ,
\end{align}
or equivalently, we modify the coupling to the background metric by $\int d^2x \sqrt{g} \alpha(J\bar{w} + \bar{J}w)$, where $w$ is the spin connection.\footnote{At this point, we might be tempted to say that these are not really anomalies because we can change them by adding local operator dependent counter-terms, and indeed this is the case.
We, however, defined our anomalies as those which we cannot remove by introducing the local c-number counter-terms. We could take the former viewpoint and then the ghost number anomaly is not an anomaly (contrary to the common use of the terminology). Of course, if we removed the ghost number anomaly in this way, then the worldsheet BRST symmetry is lost.}

Then one can immediately see that the two-point functions have the form
\begin{align}
\langle J(z) J(0) \rangle = \frac{k}{2z^2} \cr
\langle \tilde{T}(z) \tilde{T}(0) \rangle = \frac{c - 6k\alpha^2}{2z^4} \cr
\langle \tilde{T}(z) J(0) \rangle = -\frac{k\alpha}{z^3} \ .
\end{align}
This means that the $U(1)$ current under consideration realizes the impossible anomaly once we couple the theory to the background curvature through the twisted energy-momentum tensor $\tilde{T}$ rather than $T$. We note that a priori there is no preferred choices of $\alpha$ for the energy-momentum tensor in two-dimensions: they are all traceless and conserved (unlike in the other dimensions where the trace becomes non-zero by the twist). For example, in the ghost number current, the particular $\alpha$ is chosen from the other principle of the physics (e.g. worldsheet BRST symmetry). The same is true in the case of topological twist in $\hat{c}=3$ superconformal field theories. Whether we prefer the topologically twisted energy-momentum tensor to the untwisted energy-momentum tensor simply depends on the problem we would like to study.

At the same time, from the same two-point function, we see that the energy-momentum tensor has the Weyl anomaly
\begin{align}
\tilde{T}^\mu_{\mu} = -\frac{(c-6k\alpha^2)}{12} R - \frac{k\alpha}{2} \epsilon^{\mu\nu} F^A_{\mu\nu}  \label{axial}
\end{align}
under the presence of the background axial $U(1)$ curvature $F^A_{\mu\nu}$.
This gives the reciprocal relation between the impossible $U(1)$ current anomaly and the impossible Weyl anomaly. This Weyl anomaly has a tantalizing physical interpretation. Suppose we want to gauge the axial current in two-dimensional conformal field theories. The gauging would introduce the non-trivial beta functions for the gauge field strength. The Weyl anomaly \eqref{axial} indeed suggests that the $U(1)$ theta angle acquires the ``one-loop" beta function, but the point is that the coefficient is unfixed unless we specify how the theory couples to the background metric. In other words, the beta function for the $U(1)$ theta angle is completely arbitrary from the viewpoint of the flat space theory unless other principles of physics are introduced.

One caveat of this construction is that the resulting theory has the non-unitary interpretation. To see this, the twisted OPE is equivalent to the commutation relation
\begin{align}
[\tilde{L}_1, J(0)] = k\alpha  \ , \label{commut}
\end{align}
which is the manifestation of the fact that $J$ is not a conformal primary, but at the same time, the vacuum expectation values of the right hand side does not seem vanish while the left hand side does if we assume that vacuum is annihilated by $\tilde{L}_1$. Therefore, we do not have the unitary field theory interpretation of the impossible anomalies in this construction. At the same time this commutation relation makes it manifest that $J_\mu$ is not a primary operator,\footnote{This commutation relation further implies that $J_\mu$ is not a descendant either. Again this is only allowed in non-unitary conformal field theories.}  
 and this is how the anomalous conservation law \eqref{cano} is actually conformally covariant. Indeed, if we naively apply the the Wess-Zumino consistency condition for the mixed $U(1)$ transformation and the Weyl transformation, it appears to fail if we assume that Weyl transformation and the $U(1)$ transformation commute. Naively, the $U(1)$ transformation of the Weyl anomaly is zero while the Weyl transformation of the $U(1)$ anomaly is nonzero. 
However, the commutation relation \eqref{commut} actually states that they do no commute and the anomaly is consistent, which is obviously the case since we can construct examples.

Let us study three examples. The fist example is the (twisted) free fermion also known as the bc ghost system. As we already mentioned, the word-sheet ghost number current is an example of impossible anomalies.
Let us consider the free Dirac fermion
\begin{align}
 S = \int d^2x \sqrt{g} \left( \bar{\psi}_L \gamma^\mu D_\mu  \psi_L +   \bar{\psi}_R \gamma^\mu D_\mu  \psi_R \right)
\end{align}
The standard choice of the spin connection in $D_\mu = \partial_\mu + \omega_\mu \pm A_\mu$ defines the spin $1/2$ free fermion, but one can twist the fermion to have different spin connection. The twisted energy momentum tensor and $U(1)$ current are given by
\begin{align}
 T &= \frac{1}{2} \partial \bar{\psi}_L \psi_L -\frac{1}{2}\bar{\psi}_L\partial \psi_L  -\alpha \partial(\bar{\psi}_L \psi_L) \cr
 J &= -\bar{\psi}_L \psi_L 
\end{align}
The worldsheet ghost system is realized at $\alpha = \frac{3}{2}$. The fermion number is anomalous:
\begin{align}
D^\mu J_\mu = \frac{\alpha}{2} R \ .
\end{align}

The second example is the twisted free boson (also known as linear dilaton or Liouville theory):
\begin{align}
S = \int d^2x \sqrt{g} \left( \frac{1}{2} g^{\mu\nu}  \partial_\mu \phi \partial_\nu \phi + \alpha \sqrt{k} R \phi - \sqrt{k} \phi F_A^{\mu\nu} \epsilon_{\mu\nu} + \sqrt{k}\partial_\mu\phi  A_V^\mu + O(A^2)  \right) \ .
\end{align}
One may regard it as the bosonized version of the first example. Here $J= \sqrt{k}\partial \phi$ and $\bar{J} = \sqrt{k}\bar{\partial} \phi$, and note that the two-point function of the current is normalized with an extra negative sign.

The advantage of this model is that we can reproduce the anomaly from the classical analysis
\begin{align}
T^\mu_{\mu} &=  -\frac{1}{2} \alpha \sqrt{k} \Box \phi = -\frac{1}{2} \alpha^2 k R +\frac{1}{2} \alpha k \epsilon^{\mu\nu} F^A_{\mu\nu} \cr
D^\mu J_\mu^V &= \frac{1}{4}\sqrt{k} \Box \phi = \frac{1}{4}\alpha k R - \frac{1}{4} k\epsilon^{\mu\nu} F_{\mu\nu}^A \  
\end{align}
when $A^V_\mu = 0$ (up to $O(1)$ quantum correction).
In this sense, the twisted boson gives the Wess-Zumino effective action for the (impossible) anomalies. However, note that this action not only reproduces the anomalous correlation functions, but also the non-anomalous  correlation functions supported on non-coincident point, so one may regard it as a bona-fide quantum field theory with impossible anomalies.
This is a generalization of that the Liouville action or non-local Polyakov action gives the effective action for the local as well as non-local Weyl anomaly in two-dimensions.
As we will see, the discussions are more subtle in higher dimensions because local terms and non-local terms may have different origins.

Our third example is the holographic realization of the impossible anomaly.
We can realize the impossible anomaly in the holographic bulk gravity in $1+2$ dimension with the holographic topological twist. The minimal setup is to realize the three-dimensional gravity as the $SL(2,R)\times SL(2,R)$ Chern-Simons theory \cite{Witten:1988hc} and realize the $U(1)$ current sector by the $SU(1) \times U(1)$ Chern-Simons theory with the action
\begin{align}
 S = \frac{k}{4\pi} \int &\mathrm{Tr}(A \wedge dA + \frac{2}{3} A \wedge A \wedge A ) + (B \wedge dB) \cr
 &+ \mathrm{Tr}(\bar{A} \wedge d\bar{A} + \frac{2}{3} \bar{A} \wedge \bar{A} \wedge \bar{A} ) +  (\bar{B} \wedge d\bar{B}) \ .
\end{align}
To realize the topological twist, we simply replace $A_3 \to A_3 + \alpha B$ in $ A = A_1 \sigma_1 + A_2 \sigma_2 + A_3 \sigma_3$. The impossible anomaly is manifest in the boundary term in the bulk gauge transformation $B \to d\Lambda$. We also see that there is a off-diagonal two-point function of $T$ and $J$ through the kinetic term $\int A_3 dB$.

\section{Realizing impossible anomalies from semi-local term}
Another way to realize impossible anomalies is based on the semi-local terms in the correlation functions, which should be distinguished from the one in the previous section in which we had the direct implication of the impossible anomalies in non-local correlation functions supported at non-coincident points.\footnote{There has been some interest in understanding the role of semi-local terms in correlation functions in momentum space \cite{Bzowski:2013sza}\cite{Bzowski:2014qja}\cite{Bzowski:2015pba}\cite{Bzowski:2017poo}. They may be in particular important in its application to holographic cosmology (see e.g. \cite{Kawai:2014vxa} and reference therein).} 
To illustrate the idea, we again begin with the two-dimensions. 

Let us consider a two-dimensional conformal field theory with marginal coupling constants $\lambda^i$, which couples with the operator $O_i$, and the $U(1)$ current $J_\mu$. 
Let us then consider the possibility to realize the current anomaly of the following form
\begin{align}
\partial^\mu J_\mu =  \chi_{ij} \partial^\mu  \lambda^i \partial_\mu\lambda_j + \tilde{\chi}_{ij} \epsilon^{\mu\nu} \partial_\mu \lambda^i \partial_\nu \lambda^j + \cdots \ . \label{currenta}
\end{align} 
Here we have promoted the coupling constants $\lambda^i$ to be space-time dependent background scalar field.
We can easily see that they satisfy the Wess-Zumino consistency condition by assuming $\lambda_i$ are not charged under the $U(1)$ symmetry associated with the current $J_\mu$. Since they are consistent, how can we realize them?

As in the previous section, we can again propose the following no-go argument.
Suppose that $\chi_{ij}$ and $\tilde{\chi}_{ij}$ are non-zero. Then, we must be able to see it from the three-point functions of $\langle J(x) O_i(y) O_j(z) \rangle$. However, we know that the conformal invariance completely fixes  the (unregularized) three-point functions at non-coincident points in two-dimensional conformal field theory \cite{Polyakov:1970xd} as
\begin{align}
\langle J(x) O_i(y) O_j(z) \rangle = \frac{c_{ij}}{(x-y)(y-z)(z-x)(\bar{y}-\bar{z})^2} \ . \label{normal}
\end{align}
Taking the derivative with respect to $\bar{x}$, we see that it satisfies the non-anomalous Ward-Takahashi identities;
\begin{align}
\bar{\partial}_{x}\langle J(x) O_i(y) O_j(z) \rangle &=2\pi \delta^2(x-y)\frac{c_{ij}}{|y-z|^4} - 2\pi \delta^2(x-z) \frac{c_{ij}}{|y-z|^4}  \cr
& = \delta^2(x-y)q_i \langle O_i(y) O_j(z) \rangle + \delta^2(x-z) q_j \langle O_i(y) O_j(z) \rangle  \label{WT}
\end{align}
indicating  $c_{ij}$ are charges of operator $O_i$ and $O_j$ (denoted by $q_i$ and $q_j$ in the Ward-Takahashi identity). At this point, one might conclude that the anomaly of \eqref{currenta} is impossible. The semi-local term appearing in \eqref{WT} is the non-anomalous contribution, and it has nothing to do with the anomaly \eqref{currenta}.
In particular, when $O_i$ and $O_j$ are not charged, then the three-point functions at non-coincident point vanish due to the conformal invariance. How can we get anything from zero?

On the other hand, it seems that we may realize such anomalies by mimicking the free boson construction in the previous section. Let us consider the free bosonic action
\begin{align}
S = \int d^2x \left( \frac{1}{2}\partial^\mu \phi \partial_\mu \phi + \phi (\chi_{ij} \partial^\mu  \lambda^i \partial_\mu\lambda^j + \tilde{\chi}_{ij} \epsilon^{\mu\nu} \partial_\mu \lambda^i \partial_\nu \lambda^j ) \right) \ ,  
\end{align}
and we study the anomalous divergence of the vector current $J_\mu = \partial_\mu \phi$, which is conserved when $\lambda = \text{const}$. As is the case with the previous section,  one may easily see that it shows the anomaly through the classical equations of motion:
\begin{align}
\partial^\mu J_\mu = \partial^\mu \partial_\mu \phi =  \chi_{ij} \partial^\mu  \lambda^i \partial_\mu\lambda^j + \tilde{\chi}_{ij} \epsilon^{\mu\nu} \partial_\mu \lambda^i \partial_\nu \lambda^j  \ 
\end{align}
realizing the impossible anomaly. This construction shows explicitly that the Wess-Zumino consistency condition is indeed satisfied because otherwise there should be no effective field theory realization at all. This also shows that not only the anomaly equation but also the equation before computing the divergence is compatible with conformal invariance because the construction here is perfectly conformally invariant.

To see what is happening and the actual origin of the dilemma, let us compute the three-point function $\langle J(x) O_i(y) O_j(z) \rangle$ from this free boson action. In addition to the terms coming from the explicit insertion of $O_i(y)$, which vanishes when $O_i(y)$ are not charged (i.e. non-anomalous contributions), we have
\begin{align}
 \langle J(x) O_i(y) O_j(z) \rangle = &\frac{\delta}{\delta \lambda^i(y)}\frac{\delta}{\lambda^j(z)}\langle \partial \phi(x) \rangle|_{\lambda=0} \cr
 =& -\int d^2w \langle \partial \phi(x) \phi(w) \rangle \cr
&(\chi_{ij} \partial^\mu \delta^2(y-w) \partial_\mu \delta^2(z-w) + \tilde{\chi}_{ij} \epsilon^{\mu\nu} \partial_\mu \delta^2(y-w) \partial_\nu \delta^2(z-w)) \ 
\end{align}
with $\langle \phi(x) \phi(w) \rangle = \log (x-w)^2$.

In this expression, it is not difficult to check that the three-point function is indeed conformal invariant but semi-local, where the support is localized at $z=y$ with arbitrary $x$.\footnote{Under the infinitesimal special conformal transformation $x^\mu \to x^\mu + v^\mu x^2 -2(v^\rho x_\rho) x^\mu$, the delta function $\delta^{2}(x-y)$ is invariant.} We also note that the structure is intrinsically different from \eqref{normal}, and the anomalous term should have different origins than the regularization ambiguities in \eqref{normal}.

In momentum space (where we are not careful about the overall factors), we have
\begin{align}
 \langle J(k) O_i(p) O_j(q) \rangle = \delta(k+p+q) \frac{k}{|k|^2} (\chi_{ij} p^\mu q_\mu + \tilde{\chi}_{ij} \epsilon^{\mu\nu} p_\mu q_\nu) \ , \label{moma}
\end{align}
and it reproduces the anomalous divergence that we have anticipated:
\begin{align}
 \langle \bar{\partial}J(k) O_i(p) O_j(q) \rangle = \delta(k+p+q) (\chi_{ij} p^\mu q_\mu + \tilde{\chi}_{ij} \epsilon^{\mu\nu} p_\mu q_\nu) \ . \label{momb}
\end{align}
Note again this is different from the non-anomalous divergence in the momentum space
\begin{align}
\langle \bar{\partial}J(k) O_i(p) O_j(q) \rangle = \delta(k+p+q)(q_1 p^2 \log|p|^2 + q_2 q^2 \log|q|^2) \ , \label{momn}
\end{align}
which gives the non-anomalous Ward-Takahashi identity. Note that the anomalous divergence in \eqref{momb} is completely local while the non-anomalous divergence in \eqref{momn} is still semi-local.
Therefore, we may be able to construct the model of impossible anomaly but only in the semi-local terms. The existence of the anomaly is not explained by the non-anomalous part of the correlation function but is fixed by some other means.

The similar construction is available for the four-dimensional impossible anomaly of the form that we began with:
\begin{align}
\partial^\mu J_\mu = t_2F^{\mu\nu} F_{\mu\nu} \ . \label{anof}
\end{align}
Assuming that the current $J_\mu$ is a conserved primary operator, one may again argue that there is no conformally invariant  three-point function at non-coincident point that shows the structure of the impossible anomaly. This is in accord with the observation that the CP even three-point function $\langle J_\mu^a J_\nu^b J_\rho^c \rangle$ vanishes in conformal field theories unless there is a totally antisymmetric structure constant $f^{abc}$ \cite{Osborn:1993cr}.

To be more precise, the conformal invariance and the current conservation at the non-coincident point demands that the CP even part of the three-point function of the current operators at the non-coincident point must be given by a combination of the two independent terms \cite{Schreier:1971um}\cite{Freedman:1992tz}
\begin{align}
\langle J_\mu^a(x) J_\nu^b(y) J_{\rho}^c(z) \rangle = k_1^{abc} D_{\mu\nu\rho}^{\text{sym}}(x,y,z) + k_2^{abc} C_{\mu\nu\rho}^{\text{sym}}(x,y,z) \ , \label{3c}
\end{align}
where $D_{\mu\nu\rho}^{\text{sym}}(x,y,z)$ and $C_{\mu\nu\rho}^{\text{sym}}(x,y,z)$ are permutation odd tensor functions constructed out of
\begin{align}
D_{\mu\nu\rho}(x,y,z) &= \frac{1}{(x-y)^2(z-y)^2(x-z)^2} \frac{\partial}{\partial x_\mu} \frac{\partial}{\partial y_\nu} \log(x-y)^2 \frac{\partial}{\partial z_\rho} \log\left(\frac{(x-z)^2}{(y-z)^2}\right) \cr
C_{\mu\nu\rho}(x,y,z) &= \frac{1}{(x-y)^4} \frac{\partial}{\partial x_\mu} \frac{\partial}{\partial z_\alpha} \log(x-z)^2 \frac{\partial}{\partial y_\nu}\frac{\partial}{\partial z_\alpha} \log(y-z)^2 \frac{\partial}{\partial z_\rho} \log\left(\frac{(x-z)^2}{(y-z)^2}\right) 
\end{align}
by symmetrization
\begin{align}
D_{\mu\nu\rho}^{\text{sym}}(x,y,z) &= D_{\mu\nu\rho}(x,y,z) + D_{\nu\rho\mu}(y,z,x) + D_{\rho\mu\nu}(z,x,y) \cr
C_{\mu\nu\rho}^{\text{sym}}(x,y,z) & = C_{\mu\nu\rho}(x,y,z) + C_{\nu\rho\mu}(y,z,x) + C_{\rho\mu\nu} (z,x,y) \ .
\end{align}
Since the three-point function is permutation invariant, the coefficient $k_1^{abc}$ and $k_2^{abc}$ must be permutation-odd.
To fix these coefficients, we note that when we compute the divergence of \eqref{3c}, we find contact terms from $D_{\mu\nu\rho}^{\text{sym}}(x,y,z)$. These contact terms at the coincident point has the interpretation that $J^b$ is charged under $J^a$ and the symmetry group is actually non-Abelian. Then the coefficient $k_1^{abc}$ must be related to the structure constant $f^{abc}$ of the non-Abelian group through the Ward-Takahashi identity. Due to the absence of the coincident singularities in $C_{\mu\nu\rho}^{\text{sym}}(x,y,z)$, however, $k_2^{abc}$ is not fixed by the group structure. Therefore if we have more than three $U(1)$s one may have such a term in Abelian global symmetries.\footnote{One way to introduce such a term is to consider the holographic bulk theory with the three $U(1)$ gauge fields with the cubic interaction $\int d^5x\sqrt{g} F_{\mu}^{1 \ \nu} F_{\nu}^{2 \ \rho} F_{\rho}^{3 \ \mu}$.}
This is an interesting point, but since $k_2^{abc}$ is permutation odd anyway, it does not directly give rise to our anomaly because our anomaly is permutation even in $b$ and $c$. In this way, we may conclude that the conformal symmetry does not allow the impossible anomalies of the form \eqref{anof}.

 However, one may still realize the anomaly in the semi-local terms. Indeed, one may consider the free boson with the higher derivative conformal action
\begin{align}
 S = \int d^4x \left( \frac{1}{2}\phi \Box^2 \phi + B^\mu \Box \partial_\mu \phi - \phi F_{\mu\nu} F^{\mu\nu} \right) \label{effc}
\end{align}
which may be regarded as a four-dimensional dilaton theory. In order to study the  anomalous divergence of the current $J_\mu^B = \partial_\mu \Box \phi$,
 we  can compute the conformally invariant semi-local three-point function
\begin{align}
\langle J^B_\mu J_\nu J_\rho \rangle &= \int d^4w \langle \Box \partial_\mu \phi(x) \phi(w) \rangle  \left( \delta_{\nu \rho} \partial^\alpha \delta^4(w-y) \partial_\alpha \delta^4(w-z) - \partial_\nu \delta^4(w-y) \partial_\rho \delta^4(w-z) \right) \ . 
\end{align}
with the desired (impossible) anomalous divergence
\begin{align}
\langle \partial^\mu J^B_\mu J_\nu J_\rho \rangle & =  \delta_{\nu \rho} \partial^\alpha \delta^4(x-y) \partial_\alpha \delta^4(x-z) - \partial_\nu \delta^4(x-y) \partial_\rho \delta^4(x-z) \  .
\end{align}
Alternatively, in the momentum space, we have
\begin{align}
\langle \partial^\mu J^B_\mu J_\nu J_\rho \rangle = \delta(k+p+q)(p_\nu q_\rho - (pq) \delta_{\nu\rho}) \ ,
\end{align}
which is equivalent to the anomalous conservation \eqref{anof}. The structure is very similar to the two-dimensional one discussed in this section but quite different from the one discussed in the previous section. The free boson construction only gives rise to the semi-local terms and they are not directly connected to the non-local three-point functions allowed in the conformal field theories.

It is instructive to compare the results here with the one with the more conventional anomaly realized in conformally invariant Green-Schwartz mechanism. Instead of \eqref{effc}, we consider the higher derivative action
\begin{align}
 S = \int d^4x \frac{1}{2}\left(\phi \Box^2 \phi + B^\mu \Box \partial_\mu \phi - \phi \epsilon^{\mu\nu\rho\sigma} F_{\mu\nu} F_{\rho\sigma} \right)
\end{align}
One may regard it as the conformal invariant version of the Wess-Zumino action for the chiral anomaly. Alternatively, one may regard it as the conformal invariant Green-Schwartz action for the $U(1)$ current anomaly.

As before, one can compute the current three-point function as
\begin{align}
\langle J^B_\mu J_\nu J_\rho \rangle &= \int d^4w \langle \Box \partial_\mu \phi(x) \phi(w) \rangle  \left( \epsilon_{\nu\rho \alpha \beta} \partial^\alpha \delta(w-y) \partial^\beta \delta(w-z) \right) \ . 
\end{align}
In the momentum space, it is given by
\begin{align}
\langle J^B_\mu J_\nu J_\rho \rangle &=  \delta(k+p+q)\frac{k_\mu}{k^2}(\epsilon_{\mu\nu\rho\sigma}p^\rho q^\sigma) \label{semiconn}
\end{align}
or its divergence
\begin{align}
\langle \partial^\mu J^B_\mu J_\nu J_\rho \rangle = \delta(k+p+q)(\epsilon_{\mu\nu\rho\sigma}p^\rho q^\sigma) \ ,
\end{align}

Note that the semi-local term \eqref{semiconn} does not correspond to the local three-point functions that we usually obtain in the (unregularized) triangle diagram
\begin{align}
\langle J_\mu^a(x) J_\nu^b(y) J_\rho^c(z) \rangle = d^{abc} \frac{\mathrm{Tr}[\gamma_5\gamma_\mu (x_\alpha -y_\alpha)\gamma^\alpha \gamma_\nu(y_\beta-z_\beta)\gamma^\beta \gamma_\rho(z_\delta-x_\delta)\gamma^\delta)]}{(x-y)^4(y-z)^4(z-x)^4} \ , \label{usual}
\end{align}
which is known as the only conformally invariant CP-odd three-point functions of conserved current with non-zero support at non-coincident point. Also they are different from the completely local contact term ambiguities in regularizing \eqref{usual}
\begin{align}
\langle J_\mu^a(x) J_\nu^b(y) J_\rho^c(z) \rangle_{\text{amb}} = \tilde{d}^{abc} \epsilon_{\mu\nu\rho\sigma} \partial_\sigma \delta(x-y)\delta(y-z) , \label{amb}
\end{align}
that appears in the shifting momentum in the linearly divergent integral.

Nevertheless its anomalous divergence is the same as that we observe in conformal field theories. We therefore can use the semi-local terms to cancel the current anomaly as we do in the Green-Schwartz mechanism. In other words, not accepting the realization of the impossible anomalies by semi-local terms is equivalent to not allowing Green-Schwartz mechanism to cancel anomalies.

\section{Discussions on impossible Weyl anomalies}
In this paper, we have discussed two mechanisms to realize impossible anomalies. Impossible anomalies are defined such that while they satisfy the Wess-Zumino consistency conditions, they do not seem to possess the corresponding flat space conformal correlation functions. One way to circumvent the difficulty is to make the current not primary operators, and the other way to circumvent is to use the semi-local terms in the correlation functions.

There are more impossible anomalies reported in the literature, some of which will be discussed here briefly for future investigations. 
Consider the two-dimensional Weyl anomaly. With space-time dependent coupling constants $\lambda^i$, we may have
\begin{align}
T^{\mu}_{\ \mu} = cR + \eta_{ij} \partial^\mu \lambda^i \partial_\mu \lambda^j + \tilde{\eta}_{ij} \epsilon^{\mu\nu} \partial_\mu \lambda^i \partial_\nu \lambda^j + \cdots \ . 
\end{align}
The manifestation of the Weyl anomaly in the correlation functions is slightly more non-trivial than the one for the current anomaly. In \cite{Gomis:2015yaa}, they argued that $\tilde{\eta}_{ij}$ is an example of impossible anomalies. Suppose we have non-zero $\tilde{\eta}_{ij}$, then we should be able to see it from the scale anomaly in the two-point functions $\langle O_i O_j \rangle$, but there is no such terms simply because $\epsilon^{\mu\nu} p_{\mu} p_\nu = 0$, and there is no CP violating two-point functions. In contrast $\eta_{ij}$ can be directly measured in the two-point functions of $\langle O_i(p) O_j(q) \rangle = \eta_{ij} \delta(p+q) p^2 \log p^2 $. 

On the other hand, the semi-local terms that correspond to the Weyl anomaly with $\eta_{ij}$ do exist. The most convenient way to realize it to use the bosonic Liouville-like construction
\begin{align}
\int d^2x \sqrt{g} \left( \frac{1}{2}\partial_\mu \phi \partial^\mu \phi + R \phi + \phi (\tilde{\eta}_{ij} \epsilon^{\mu\nu} \partial_\mu \lambda^i \partial_\nu \lambda^j) \right) \ . \label{bos}
\end{align}
It is more non-trivial to see how the induced Weyl anomaly affects the correlation functions than the case in the $U(1)$ current anomaly. Instead of computing the insertion of the trace of the energy-momentum tensor directly, let us compute the three-point functions of $\langle T(k) O_i(p) O_j(q) \rangle$ from \eqref{bos}. In the momentum space it is given by the semi-local term
\begin{align}
\langle T(k) O_i(p) O_j(q) \rangle = \delta(k+q+p) \frac{k}{\bar{k}} \epsilon_{\mu\nu}p^\mu q^\nu \ ,
\end{align}
so its divergence gives the contact term
\begin{align}
 \langle \bar{\partial}  T(k) O_i(p) O_j(q) \rangle = \delta(k+q+p) {k} \epsilon_{\mu\nu}p^\mu q^\nu \ .
\end{align}
This term does not come from the semi-local terms in the ordinary Ward-Takahashi identity of the energy-momentum tensor conservation, so it must be cancelled from the insertion of $\partial T^\mu_{\ \mu}$ to avoid the gravitational anomaly, which in turn means that the trace of the energy-momentum tensor has the (semi-)local terms in the flat space-time limit.
\begin{align}
 \langle T^{\mu}_{\ \mu}(k) O_i(p) O_j(q) \rangle = \delta(k+q+p) \epsilon_{\mu\nu}p^\mu q^\nu
\end{align}
This is not inconsistent with the dilatation Ward-Takahashi identity because 
\begin{align}
\delta_{\text{dilataion}}  \langle O_i(p) O_j(q) \rangle  =  \lim_{k_\mu \to 0}\langle T^{\mu}_{\ \mu}(k) O_i(p) O_j(q) \rangle =  \delta(q+p) \epsilon_{\mu\nu}p^\mu q^\nu = 0 
\end{align}
and as we saw, there is no CP odd term in two-point functions, and the CP odd term in the left hand side is zero from the beginning.
 Thus the Weyl anomaly here does not correspond to the dilatation anomaly in the two-point function.

The similar construction is possible for the four-dimensional Weyl anomaly. There has been some debates over whether the CP-odd Pontryagin term can appear in the four-dimensional Weyl anomaly \cite{Nakayama:2012gu}\cite{Bonora:2014qla}\cite{Bonora:2015nqa}. It satisfies the Wess-Zumino consistency condition, but there is no conformally invariant CP-odd three-point functions of $\langle T_{\mu\nu} T_{\rho\sigma} T_{\alpha \beta} \rangle$ supported at non-coincident points. Therefore it is an example of impossible anomalies in our terminology. 

Our construction, however, suggests that at least the semi-local correlation functions including the Pontryagin Weyl anomaly exist. 
To see this, we consider the Riegert-type effective action for the four-dimensional Pontryagin Weyl anomaly given by 
\begin{align}
S = \int d^4x \sqrt{g} \left( \frac{1}{2} \phi \Delta_4 \phi - Q\mathcal{Q}\phi - \phi \epsilon^{\alpha\beta \gamma \delta} R_{\alpha\beta \mu\nu} R^{\mu\nu}_{\ \ \gamma \delta}  \right) \ ,
\end{align}
where $\Delta_4 = \Box^2 + 2G_{\mu\nu} D^\mu D^\nu + \frac{1}{3}(D^\mu R)D_\mu + \frac{1}{3} R \Box $ is the Fradkin-Tseytlin-Riegert-Paneitz operator \cite{Fradkin:1982xc}\cite{Fradkin:1981jc}\cite{Riegert:1984kt}\cite{PA} and $\mathcal{Q} = -\frac{1}{6} \Box R - \frac{1}{2}R^{\mu\nu} R_{\mu\nu} + \frac{1}{6} R^2$ is the so-called Q-curvature \cite{Q}. The classical equations of motion gives the Weyl anomaly
\begin{align}
T^{\mu}_{\ \mu} = Q\Delta_4 \phi  = Q^2 \mathcal{Q} + Q\epsilon^{\alpha\beta \gamma \delta} R_{\alpha\beta \mu\nu} R^{\mu\nu}_{\ \ \gamma \delta} \ 
\end{align}
due to the nice conformal properties of Q-curvature:
\begin{align}
\mathcal{Q} \to e^{-4\sigma}(\mathcal{Q} + \Delta_4 \sigma) \ .
\end{align}
This effective action does reproduce 
\begin{align}
\langle T^{\mu}_{\ \mu}(x) T_{\sigma \rho}(y) T_{\alpha\beta}(z) \rangle = Q\epsilon_{\sigma \alpha\epsilon \kappa}[\partial_\beta\partial_\rho - \partial^2 \delta_{\beta\rho})(\partial^\epsilon \delta(x-y)\partial^\kappa\delta(x-z)] + \text{sym} 
\end{align}
reported in \cite{Nakayama:2012gu}\cite{Bonora:2014qla} in its CP odd part.

The realization of the Weyl anomalies from semi-local terms need further comments. There has been another debate if the CP even part of the Weyl anomaly can reproduce the flat space conformal three-point functions in this Riegert-type action \cite{Erdmenger:1996yc}\cite{Coriano:2017mux}. Since we do not have the non-coincident three-point functions with CP violation anyway (see e.g. \cite{Costa:2011mg}\cite{Zhiboedov:2012bm}), we are not concerned about this point, but further studies would clarify the role of our effective action in relation to the one-loop computations in \cite{Bonora:2014qla}. By the same token, the two-point functions of the energy-momentum tensor at the non-coincident point do not depend on $Q$ as observed in \cite{Levy:2018bdc} while the Weyl anomaly does. This is not a contradiction because the Weyl anomaly is realized by semi-local terms in this model, but we may miss the connection between the energy-momentum tensor central charge $C_T$ and the Weyl anomaly.\footnote{In particular, in the Riegert-type action, we can further change the Weyl anomaly coefficient $c$ by adding $\int d^4x \sqrt{g} \phi \mathrm{Weyl}^2$ as we like while keeping $C_T$ fixed.}

The existence of the conformally invariant semi-local terms to realize the impossible Weyl anomaly means that such an anomaly is indeed consistent, which should be contrasted with the inconsistent Weyl anomaly such as $R^2$, but we should realize that the above effective realization requires the non-unitary field theories in four-dimensions because of the higher derivative kinetic term. Thus, even if such amplitudes can be found in actual conformal field theories in a certain regularization, we might have to worry about the unitarity of such theories. With this regard, it would be more satisfactory to see if there are more intrinsic problems in these impossible anomalies from the direct studies of correlation functions rather than from a particular realization.

\section*{Acknowledgements}
The author would like to thank Z.~Komargodski and J.~Gomis for the correspondence on the impossibilities of Pontryagin Weyl anomaly in conformal field theories. He also would like to thank K.~Skenderis for the correspondence on the current three-point functions in conformal field theories.
This work is in part supported by JSPS KAKENHI Grant Number 17K14301.

\end{document}